\documentclass[10pt,aps,prb,twocolumn,superscriptaddress]{revtex4-2}
\usepackage{amsmath}
\usepackage{mathtools} 
\usepackage{amssymb}
\usepackage{amsthm}
\usepackage{bm} 
\usepackage{upgreek} 
\usepackage{xcolor}
\usepackage{graphicx}
\usepackage{epstopdf}
\usepackage{ulem}
\usepackage[colorlinks,linkcolor=blue,anchorcolor=blue,citecolor=blue,urlcolor=blue]{hyperref}
\usepackage{braket}
\usepackage{orcidlink}
\usepackage{siunitx}
\usepackage[english]{babel}
\usepackage{svg}
\usepackage{soul}
\usepackage{diagbox}
\usepackage{bbm}
\usepackage{dsfont}
\sethlcolor{yellow}

\newcommand{\nn}{{\nonumber}}

\newcommand{\1}[1]{\mathrm{#1}}

\newcommand{\LNO}{La$_3$Ni$_2$O$_{7}$}

\newcommand{\LAO}{LaAlO$_{3}$}

\begin{document}
\title{Tunable superconductivity and spin density wave in \LNO/\LAO \ thin films}

\author{Yu-Han Cao\orcidlink{0000-0002-2921-6302}} \thanks{These authors contributed equally to this work.}
\affiliation{National Laboratory of Solid State Microstructures $\&$ School of Physics, Nanjing University, Nanjing 210093, China}

\author{Kai-Yue Jiang\orcidlink{0009-0007-0395-9662}} \thanks{These authors contributed equally to this work.}
\affiliation{National Laboratory of Solid State Microstructures $\&$ School of Physics, Nanjing University, Nanjing 210093, China}

\author{Hong-Yan Lu\orcidlink{0000-0003-4715-7489}} \email[Contact author: ]{hylu@qfnu.edu.cn}
\affiliation{School of Physics and Physical Engineering, Qufu Normal University, Qufu 273165, China}

\author{Da Wang\orcidlink{0000-0003-1214-6237}} \email[Contact author: ]{dawang@nju.edu.cn}
\affiliation{National Laboratory of Solid State Microstructures $\&$ School of Physics, Nanjing University, Nanjing 210093, China}
\affiliation{Collaborative Innovation Center of Advanced Microstructures, Nanjing University, Nanjing 210093, China}

\author{Qiang-Hua Wang\orcidlink{0000-0003-2329-0306}} \email[Contact author: ]{qhwang@nju.edu.cn}
\affiliation{National Laboratory of Solid State Microstructures $\&$ School of Physics, Nanjing University, Nanjing 210093, China}
\affiliation{Collaborative Innovation Center of Advanced Microstructures, Nanjing University, Nanjing 210093, China}

\begin{abstract}
Recently, \LNO \ thin film on the \LAO \ substrate is shown to be superconducting, while the bulk \LNO \ with the same in-plane lattice constant under pressure does not superconduct. This difference suggests the interlayer distance $d_{\rm Ni-Ni}$ is crucial to control superconductivity, and its variation under pressure may tune the ground state sensitively. We investigate systematically the \LNO/\LAO \ thin films in a reasonable range of $d_{\rm Ni-Ni}$, by a combination of the first-principle calculations and the singular-mode functional renormalization group. For smaller (larger) $d_{\rm Ni-Ni}$, the ground state is a C-type (G-type) spin density wave with spins coupled ferromagnetically (antiferromagnetically) across the two layers. Between the two phases, $s_\pm$-wave superconductivity emerges with dominant pairings between nickel $3d_{3z^2-r^2}$ orbitals. The results explain the experimental superconductivity in the thin film under ambient pressure, and predict that the applied pressure will decrease the superconducting transition temperature, until the system enters the C-type spin density wave. Experimental verification would provide profound insights into the nature of electron correlations in this system, since the C-type spin density wave is achieved most naturally in the itinerant picture, while it would be hard in the local moment picture where spins are always coupled antiferromagnetically across the layers. 
\end{abstract}
\maketitle

{\it Introduction}.  
The discovery of high-temperature superconductivity in Ruddlesden-Popper (RP) phase multilayer nickelates — specifically in \LNO \ with a transition temperature $T_c$ reaching the liquid nitrogen regime under high pressure \cite{2023Nature} — raises intense experimental \cite{exp-6,exp-7,exp-8,exp-9,exp-10,exp-11,exp-13,exp-16,exp-28,exp-37,exp-38,exp-41,exp-43,exp-44,exp-48,exp-49,exp-50,exp-51,exp-53,exp-54,exp-56,exp-57,exp-61,exp-68,exp-69,327pressure,exp-70,exp-71,exp-72,exp-73,exp-74,exp-75,exp-76,exp-77,exp-78,exp-79,exp-80,exp-81,exp-82,exp-83,r-1,r-2} and theoretical \cite{t-1,t-2,t-5,t-6,t-7,t-11,t-12,t-14,t-15,t-16,t-17,t-18,t-20,t-21,t-22,t-23,t-24,t-27,t-28,t-36,t-39,t-55,t-58,t-62,t-63,t-64,t-67,t-73,t-76,t-77,t-78,t-82,liuyq_2025_VMC327,r-1,r-2,LNO_pressure_PRL_2025} research interest. This breakthrough establishes nickelates as a third family for studying high-$T_c$ superconductivity, beyond cuprate and iron-based superconductors.
Although most theoretical studies predict $s_\pm$-wave pairing from the very beginning \cite{t-2}, the proposed  superconducting mechanisms are roughly divided into two categories. One is based on local moment of the nickel $3d_{3z^2-r^2}$ orbitals in the strong coupling limit. The other is based on the itinerant picture where spin or charge fluctuations mediate the superconductivity. The observed monotonic decrease of $T_c$ with increasing hydrostatic pressure in bulk nickelates \cite{327pressure} is naturally explained in the itinerant picture \cite{LNO_pressure_PRL_2025}, where the increased bandwidth weakens the spin fluctuations, but the interpretation in the local moment picture is more intricate \cite{ft-8}. 

Recently, RP nickelate thin films on SrLaAlO$_{4}$ (SLAO) substrates \cite{fe-1,fe-2,fe-5,fe-7,fe-8,fe-12,fe-13,fe-14,fe-15,fe-16,fe-22,me-5} are explored extensively for superconductivity under ambient pressure. The in-plane compressive strain is about 2\%, with the in-plane (out-of-plane) lattice constant $a$ ($c$) similar to (much larger than) that in the bulk material under 14~GPa. The chemical pressure in the thin film is therefore a very anisotropic analogue of the bulk case. On the other hand, Sr diffusion into \LNO \ causes an effective doping of $0.2$ holes per nickel in \LNO. Angle-resolved photo-emission and scanning tunneling microscopy results support $s$-wave pairing symmetry in (La,Pr)$_3$Ni$_2$O$_{7}$ films \cite{fe-13,fe-4}. Theoretically \cite{Cao_SCPMA_2025}, it is predicted that $T_c$ can be enhanced further by (i) a reduction of the charge transfer toward the nominal electron filling in the nickel layers, and (ii) an increase of the compressive strain (namely, smaller $a$) or increase of $c$. The pathway (i) turns out to be consistent with the much higher $T_c$ in the La$_3$Ni$_2$O$_7$/La$_2$NiO$_4$/SrLaAlO$_{4}$ thin film \cite{fe-28}, where the La$_2$NiO$_4$ buffer layer could reduce the charge transfer into the \LNO\ layers. The pathway (ii) is consistent with the lower $T_c\sim 12$K in La$_2$PrNi$_2$O$_{7}$/LaAlO$_{3}$ (LAO) thin films, where the in-plane compressive strain exerted by the LAO substrate is about 1.2\% \cite{fe-27}, much lower than that induced by the SLAO substrate.  
More interestingly, we observe that the bulk \LNO \ would not superconduct if its in-plane $a$ is the same as that in La$_2$PrNi$_2$O$_{7}$/LaAlO$_{3}$. 
This comparison indicates that the variation of the out-of-plane lattice constant $c$ in the thin films is not only sensitive to superconductivity, but can even change the ground state. Note that $c$ is not well defined in a double-layer thin film, but we believe it is positively correlated with the interlayer distance $d_{\rm Ni-Ni}$. This parameter can be tuned by applying pressure on the thin film, without significant change in the in-plane $a$, as discussed in a recent experiment for the \LNO/SrLaAlO$_{4}$ thin film \cite{fe-16}. 

Motivated by the above considerations, we investigate the \LNO/LaAlO$_{3}$ thin films systematically for a reasonable range of interlayer nickel-nickel distance $d_{\rm Ni-Ni}$, assuming that the in-plane lattice constant $a$ is basically pinned by the substrate. The electronic structures are obtained by the first-principle calculations, followed by tight-binding fit for the bilayer \LNO. We then determine the ground state arising from electronic correlation effects by the singular-mode functional renormalization group. For smaller (or larger) $d_{\rm Ni-Ni}$, the ground state is a C-type (or G-type) spin density wave (SDW) with spins coupled ferromagnetically (or antiferromagnetically) across the two layers. Between the two phases, the ground state is an $s_\pm$-wave superconductor with Cooper pairs mainly composed of nickel $3d_{3z^2-r^2}$ orbitals. We predict that the applied pressure will decrease the superconducting $T_c$ until the system enters the C-type SDW. This could be tested by further experiments, and would shed profound insights into the nature of electron correlations in \LNO, since the C-type SDW is achieved most naturally in the itinerant picture, while it would be hard in the local moment picture where spins are always coupled antiferromagnetically across the layers.


\begin{figure*}
    \includegraphics[width=1\linewidth]{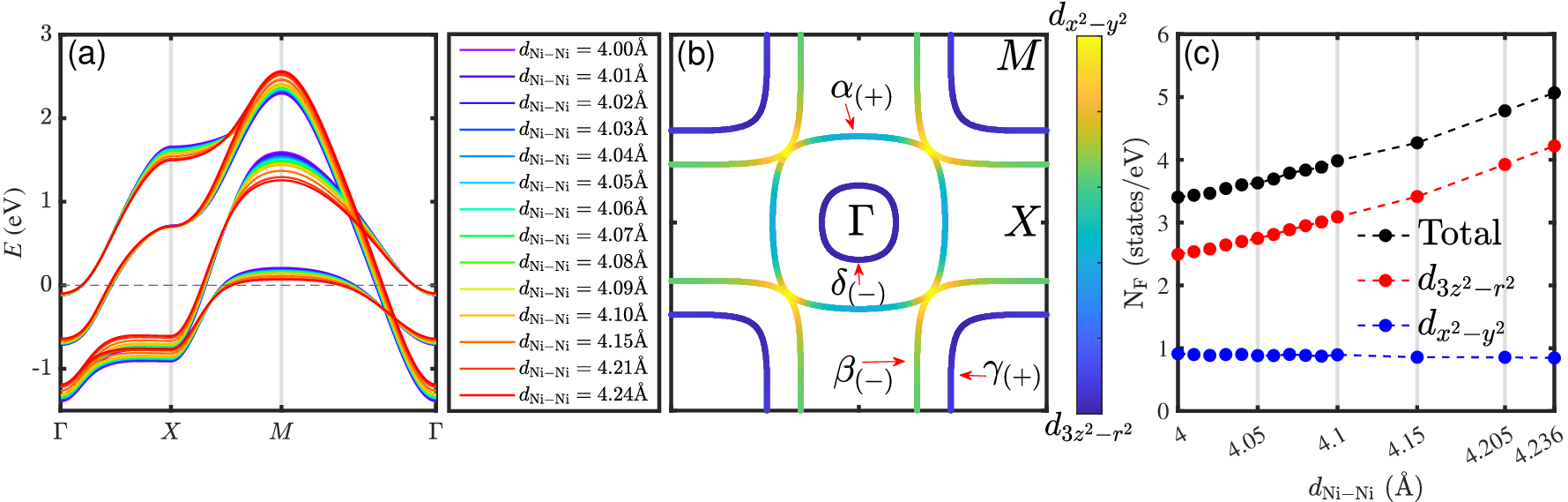}
	\caption{ (a) Tight-binding band dispersions of the \LNO \ film on LAO substrates with different interlayer Ni-Ni distance $d_\1{Ni-Ni}$. (b) The Fermi surfaces for $d_\1{Ni-Ni}=4.05$\r{A} with color-scaled orbital weight. (c) The density of states at the Fermi level ($N_F$) of total, $z$ and $x$ orbitals, respectively.}
    \label{band_FS}
\end{figure*}

{\it Model and method}.
We perform density functional theory (DFT) calculations on the \LNO \ thin films with the in-plane lattice constant $a=3.787$~\AA, corresponding to an in-plane compressive strain of 1.2\%, for a series of interlayer nickel-nickel distance $d_\1{Ni-Ni}$. The appropriate range of this parameter is determined in Sec.~IB, Table~S1, and Fig.~S1 of the Supplemental Materials (SM \cite{SM}).  The band structures are first calculated using the Vienna {\it ab initio} Simulation Package (VASP)~\cite{VASP}. In this work, the effects of Pr doping are not explicitly taken into account, since the primary role of Pr substitution is found to suppress oxygen vacancies and enhance the superconducting volume fraction \cite{exp-37,fe-1,fe-3,fe-27} but not significantly modify the low-energy orbital characters near the Fermi level \cite{t-76}. 
After obtaining the band structures, the maximally localized Wannier functions (MLWF) \cite{Wannier1,Wannier2}, as implemented in the Wannier90 code~\cite{Wannier3}, are employed to construct a bilayer tight-binding model with the nickel $3d_{x^2-y^2}$ (denoted as $x$) and $3d_{3z^2-r^2}$ (denoted as $z$) orbitals
\begin{equation}
H_0=\sum_{i \delta, a b, \sigma} t_\delta^{a b} c_{i a \sigma}^{\dagger} c_{i+\delta b \sigma}
+\sum_{i a \sigma} \varepsilon_a c_{i a \sigma}^{\dagger} c_{i a \sigma},
\end{equation}
where $t_\delta^{a b}$ denotes the hopping matrix element between the $a$ orbital on site $i$ and the $b$ orbital on site $i+\delta$, $\sigma$ is the spin index, and $\varepsilon_a$ is the on-site energy of the $a$ orbital. The Fermi energy is set to zero, corresponding to $1.5$ electrons per Ni atom. 
Details of the first-principle calculations, the orbital-projected DFT bands for the $x$ and $z$ orbitals of Ni, as well as comparisons between the DFT and Wannier band structures, are presented in Sec.~IA, Fig.~S2, and Fig.~S3 of the SM \cite{SM}, respectively. All tight-binding parameters for various values of $d_\1{Ni-Ni}$ are summarized in Fig.~S4, Table~S2, and Table~S3 of the SM.

Fig.~\ref{band_FS}(a) shows the tight-binding band dispersion of the \LNO \ films. Note that as $d_{\1{Ni-Ni}}$ increases, the bandwidth $2|t_{(00\frac12)}^{z z}|$ of the $z$ orbital at the $M$ point (lower two bands) decreases. Concurrently, both the bonding and antibonding bands of the $z$ orbital, $\varepsilon_{z} - 4t^{zz}_{(100)} \pm t^{zz}_{(00\frac12)}$, decrease towards the Fermi level due to the pronounced decrease of $\varepsilon_{z} - 4t^{zz}_{(100)}$. 
In contrast, the two nearly degenerate bonding and antibonding energies of the $x$ orbitals (upper two bands) rise owing to a significant increase in $\varepsilon_{x}$, leading to an enhancement of total bandwidth. 
Fig.~\ref{band_FS}(b) displays the Fermi surfaces for $d_\1{Ni-Ni}=4.05$~\r{A}. There are two pockets around the $\Gamma$ point, labeled as $\alpha_{(+)}$ and $\delta_{(-)}$, with the subscript exhibiting the parity of the Bloch state under the mirror operation that interchanges the two layers. (Note the substrate breaks the mirror symmetry geometrically, but we assume the mirror symmetry remains in the electronic structure to a good approximation.) Around the $M$ point, there are two pockets labeled by $\beta_{(-)}$ and $\gamma_{(+)}$, respectively. 
In the following, the parity is omitted for brevity unless specified otherwise. 
As $d_{\1{Ni-Ni}}$ increases, the $\alpha$ and $\delta$ pockets barely change, while the $\beta$/$\gamma$ pocket is visibly expanded/contracted.
Unlike the bulk \LNO, there is an additional $\delta$ pocket around the $\Gamma$ point in thin films on both SLAO \cite{Cao_SCPMA_2025} and LAO substrates for $n=1.5$ $E_g$-electrons per Ni atom in the two-orbital model. Similar to the case on the SLAO substrate \cite{Cao_SCPMA_2025}, an increase of $d_{\1{Ni-Ni}}$ enhances the density of states at the Fermi level ($N_F$), as shown in Fig.~\ref{band_FS}(c). In particular, as $d_{\mathrm{Ni-Ni}}$ increases, $N_F$ for the $z$ orbital increases more significantly, whereas that for the $x$ orbital remains almost unchanged. 

We study the effect of electronic correlations by including the atomic multi-orbital Coulomb interactions, 
\begin{align}
H_I= & \sum_{i, a<b, \sigma \sigma^{\prime}}\left(U^{\prime} n_{i a \sigma} n_{i b \sigma^{\prime}}+J_H c_{i a \sigma}^{\dagger} c_{i b \sigma} c_{i b \sigma^{\prime}}^{\dagger} c_{i a \sigma^{\prime}}\right) \nn\\
& +\sum_{i a} U n_{i a \uparrow} n_{i a \downarrow}+\sum_{i, a \neq b} J_P c_{i a \uparrow}^{\dagger} c_{i a \downarrow}^{\dagger} c_{i b \downarrow} c_{i b \uparrow} ,
\end{align}
where $U$ and $U'$ are the intra-orbital and inter-orbital Coulomb repulsion, $J_H$ is the Hund's coupling, and $J_P$ is the pair hopping interaction. We assume the Kanamori relations $U=U'+2J_H$ and $J_H=J_P$ \cite{KanamoriRelations}, and we take the two independent interaction parameters $U = 3$~eV and $J_H = 0.3$~eV. We have checked that a slight change in these parameters does not change the conclusions qualitatively. We perform singular-mode functional renormalization group (SM-FRG) calculations to investigate the correlation effects. It traces the evolution of the one-particle irreducible four-point interaction vertex function $\Gamma_{\Lambda}$ (starting from the bare interactions) versus a decreasing infrared cutoff energy scale $\Lambda$. In SM-FRG, the same vertex function $\Gamma_\Lambda$ is reformulated as scattering matrices between fermion bilinears in the SC, SDW, and CDW channels, with the channel overlaps taken into account consistently. In this manner, the competing orders are treated on equal footing. The divergence of the leading negative singular value $S_{\Lambda}$ of the scattering matrices, out of all momenta and all channels, signals an emerging long-range order, described by the leading eigenmode (and its momentum), and the divergence scale $\Lambda_c$ is representative of the transition temperature $T_c$. Further technical details can be found in Refs.~\cite{Wang_PRB_2012, Wang_PRB_2013, Tang_PRB_2019, Yangqg_prb_2022, t-2, Yangqg_prb_2024_4310, Yangqg_prb_2024_CBO, LNO_pressure_PRL_2025,ly_O}, and also in the SM \cite{SM} for self-completeness.

\begin{figure}[b]
	\includegraphics[width=\linewidth]{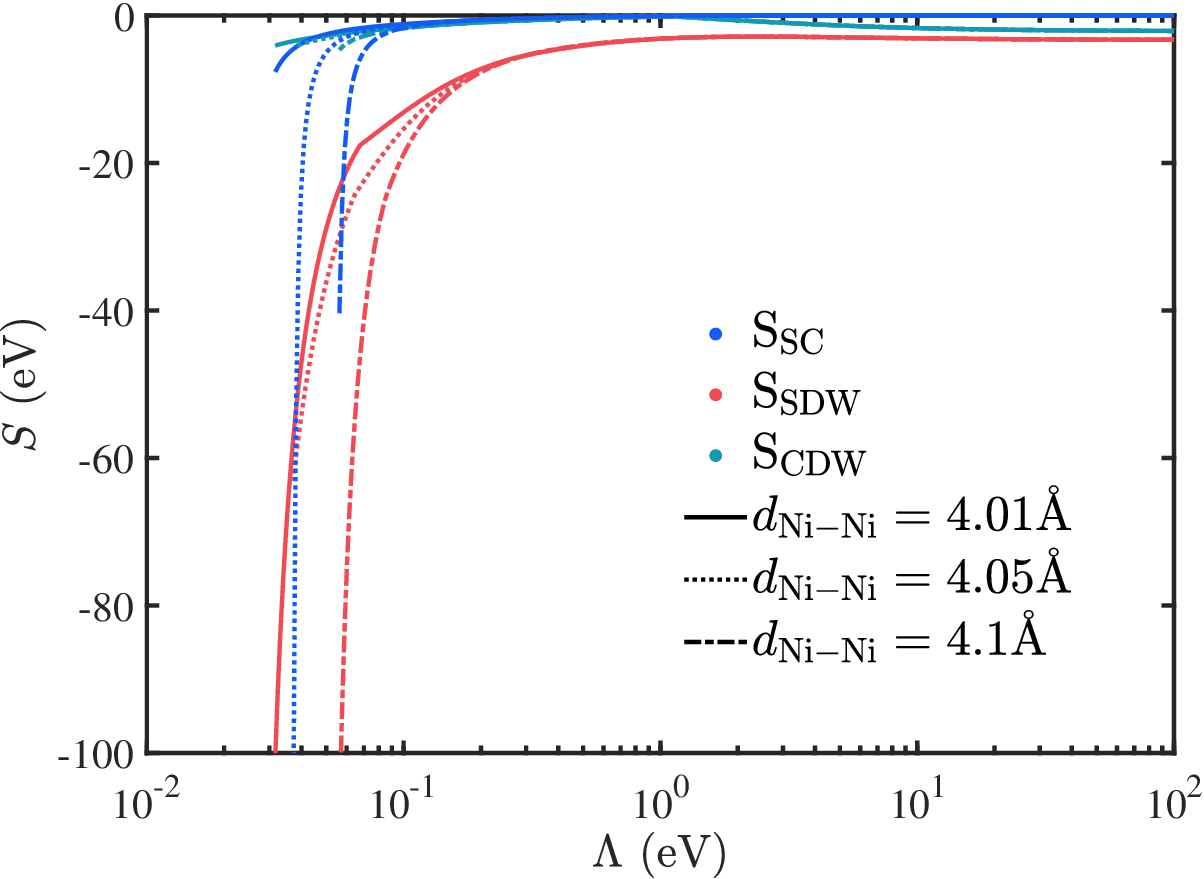}
	\caption{SM-FRG flows of the leading singular value $S$ in the SC, SDW and CDW channels versus the energy scale $\Lambda$ for $d_{\1{Ni-Ni}}=4.01$~\r{A}, 4.05~\r{A} and 4.1~\r{A}, respectively.}
	\label{flow}
\end{figure}

\begin{figure*}
    \includegraphics[width=0.8\linewidth]{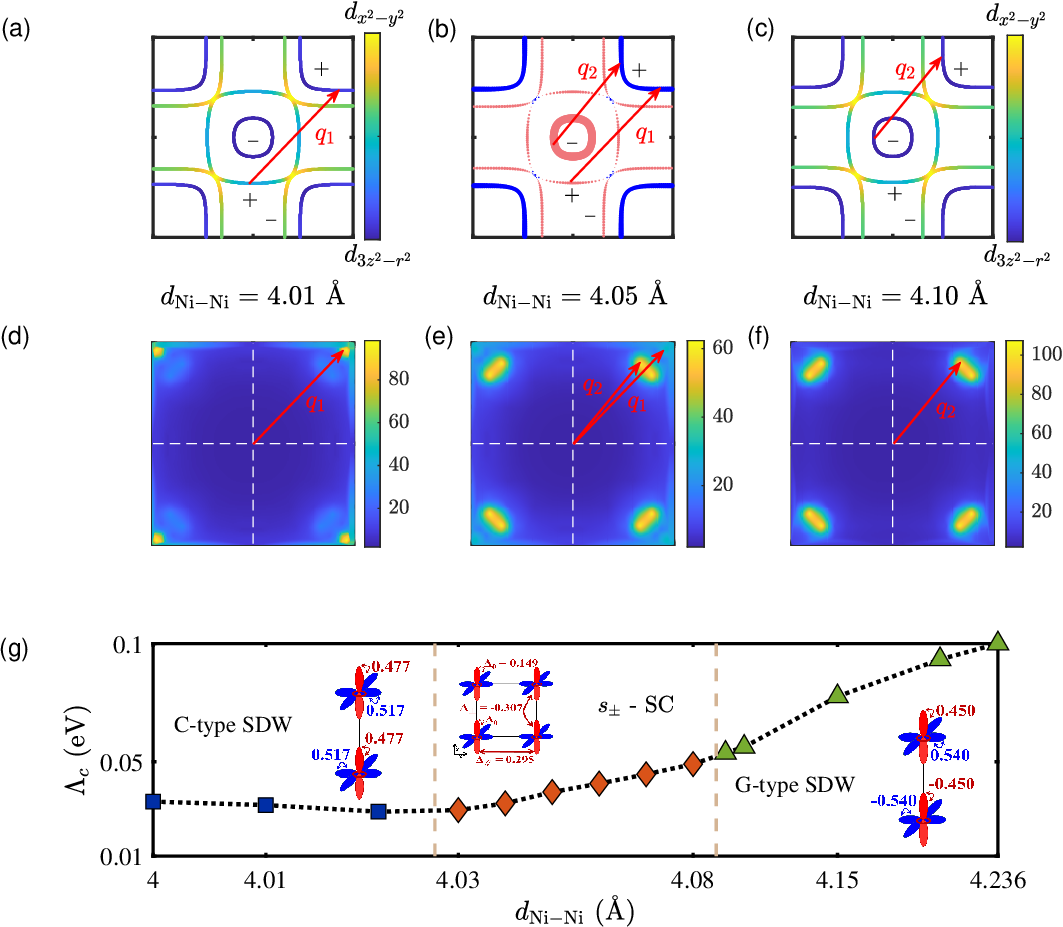}
	\caption{ Phase diagram of \LNO \ film on LAO substrates obtained by SM-FRG. (a) and (c) show Fermi surfaces for $d_{\1{Ni-Ni}}=4.01, 4.1$\AA, respectively, with color-scaled orbital weight. The arrows indicate two dominant spin scattering vectors $\0q_1$ and $\0q_2$. The mirror parities of the pockets connected by the scattering momenta are also indicated explicitly. (b) shows the SC gap function on the Fermi pockets for $d_{\1{Ni-Ni}}=4.05$\AA, where the red/blue colors indicate the positive/negative sign, and the width of the line indicates the magnitude of the gap. (d-f) show the momentum dependent SDW interaction for $d_{\1{Ni-Ni}}=4.01, 4.05, 4.1$\AA, respectively, with the arrows indicating the peak momentum $\0q_1$ and $\0q_2$. (g) shows the transition temperature  of the ordered states versus $d_{\1{Ni-Ni}}$. Left, middle, and right insets of (g) are the real-space structure of the leading eigenmodes, respectively. The numbers in left and right inset show the relative spin configurations within a unitcell, and in middle inset shows the dominant pairing components between the $z$-orbitals: onsite pairing $\Delta_0$, interlayer pairing $\Delta_\perp$, and intralayer nearest-neighbor pairing $\Delta_{\parallel}$.}
	\label{phase}
\end{figure*}

{\it Results and discussions.}
A bird's-eye view of the FRG flow is shown in Fig.~\ref{flow}, where we present the most negative singular value $S$ in the SC, SDW and CDW channels versus the energy scale $\Lambda$ for $d_{\1{Ni-Ni}}=4.01$~\r{A}, 4.05~\r{A} and 4.1~\r{A}, respectively. These eigenvalues, or mode-mode interactions, barely change at high energy scales. Because  of the local repulsive Coulomb interactions, the SDW channel is the strongest initially. As $\Lambda$ decreases, the CDW channel is suppressed by the screening effect. When $\Lambda$ drops below the bandwidth $\sim1$~eV, the SDW channel is enhanced, and meanwhile, the SC and CDW channels are also enhanced. In this sense, the spin fluctuations promote the CDW and SC through channel mixing. At low energy scales, one of the channels would diverge, implying an emerging order. The ordering channel varies with $d_{\1{Ni-Ni}}$, as we showcase below. 

The SDW state is obtained at a lower value, $d_{\rm Ni-Ni}=4.01$~\AA. Fig.~\ref{phase}(a) shows the Fermi pockets, and Fig.~\ref{phase}(d) shows the leading SDW interaction (out of all eigenmodes) as a function of in-plane momentum. The peak momentum $\0q_1$ is close to $(\pi,\pi)$, therefore it is antiferromagnetic within the plane. From the leading SDW scattering mode, we can further determine the spin structure. We find it is dominated by onsite spins, and they are arranged as in Fig.~\ref{phase}(g) (left inset) within the double-layer unitcell. The spins are parallel locally within the two orbitals due to Hund's coupling, but surprisingly, they are parallel across the layers either. We dub this as C-type SDW. We note that the interlayer ferromagnetic spin correlation would be difficult to understand in the local moment picture, where the $z$-orbitals always develop antiferromagnetic superexchange across the layers. In the itinerant picture we are working, the C-type SDW can be understood naturally as follows \cite{ft-3}. We observe that the SDW momentum $\0q_1$ connects the $\alpha_{(+)}$ and $\gamma_{(+)}$ pockets, both of even mirror parity, see Fig.~\ref{phase}(a). 
The leading SDW eigenmode can be written as $\sum_{k}\psi_{k+q_1\uparrow}^\dagger O \psi_{k\downarrow}$, where $\psi_k$ is the spinor fermion field at momentum $k$, and $O$ is a diagonal matrix in the orbital and layer basis characterizing the eigenmode. By mirror symmetry, the Bloch states carry definite mirror parity. The even (odd) parity state is $|\pm\rangle =(1,\pm 1)/\sqrt{2}$ in the layer basis. On the other hand, the diagonal matrix $O$ can be decomposed into parity conserving and parity reversing components. They correspond to the Pauli matrix $\sigma_{0,3}$ in the layer basis, respectively. If the spin scattering involves electron states of equal parity, as in the present case, only the parity conserving $\sigma_0$-component of the matrix $O$ survives, since $\langle +|\sigma_3|+\rangle=0$. This means the spins are parallel across the layers, and the SDW is of C-type.     

The SC state is obtained at an intermediate value, $d_{\rm Ni-Ni}=4.05$~\AA. Fig.~\ref{phase}(b) shows the superconducting gap function projected onto the Fermi surfaces. The function is invariant under four-fold rotation, and changes sign from $(\alpha,\beta,\delta)$ to $\gamma$ pocket. This shows the $s_\pm$-wave pairing symmetry. Fig.~\ref{phase}(g) (middle inset) shows the real-space components of the pairing eigenmode. We observe that the dominant component is the interlayer pairing between the $z$-orbitals, $\Delta_\perp$, and the subleading one is the intralayer nearest-neighbor pairing $\Delta_\parallel$, also between the $z$-orbitals, which is even comparable to $\Delta_\perp$. (The components involving $x$-orbitals are negligibly small.)
Keeping the z-orbital components only, the pairing matrix can be written in the layer basis as
\begin{align*}  
\begin{bmatrix}
\Delta_0+2\Delta_\parallel(\cos k_x+\cos k_y) & \Delta_\perp \\
\Delta_\perp & \Delta_0+2\Delta_\parallel(\cos k_x+\cos k_y)
\end{bmatrix},
\end{align*}
where $\Delta_0$ is the onsite intra-orbital pairing. 
Projected onto the band basis with definite parity $\nu=\pm 1$, the gap function is
\begin{align*}
\Delta_\nu (\0k)=\rho_{\0k\nu}(\Delta_0+2\Delta_\parallel(\cos k_x+\cos k_y)+\nu \Delta_\perp),
\end{align*}
where $\rho_{\0k\nu}$ is the $z$-orbital weight of the Bloch state $|\0k\nu\rangle$. 
We find this function well reproduces the sign structure in Fig.~\ref{phase}(b). (It also explains the small gap values on the $\alpha$ and $\beta$ pockets, where the $z$-orbital weight is small.) 
We note that the situation here is quite different from that in the bulk case and in thin films on SLAO \cite{Cao_SCPMA_2025}, where $|\Delta_\parallel|\ll |\Delta_0|<|\Delta_\perp|$ such that the relative sign of the gap function is roughly determined by the mirror parity alone \cite{t-2,Cao_SCPMA_2025,LNO_pressure_PRL_2025}. The difference in the sign structure may affect the properties of the superconducting states, such as the spin susceptibilities and the quasiparticle scattering interference against impurities, which we leave for future investigations.   

Another type of SDW state is obtained at a larger value, $d_{\rm Ni-Ni}=4.1$~\AA. Fig.~\ref{phase}(c) shows the Fermi pockets, and Fig.~\ref{phase}(f) shows the leading SDW interaction as a function of in-plane momentum. The peak momentum $\0q_2$ is close to $(\pi,\pi)$ but slightly deviates from the diagonal. So the SDW is also antiferromagnetic within the plane. The leading SDW mode is shown in Fig.~\ref{phase}(g) (right inset) within the unitcell. The spins are parallel within the two orbitals (again due to the local Hund's coupling), but are antiparallel across the layers. We dub this as G-type SDW. The difference from the C-type SDW is because the SDW momentum $\0q_2$ now connects Fermi pockets of opposite mirror parity, as shown in Fig.~\ref{phase}(c). The same argument as above immediately tells us that now the layer-odd spin structure is consistent with the parity flip in the spin scattering at momentum $\0q_2$, since the layer-even component $\langle+|\sigma_0|-\rangle$ vanishes. 

Given the very different spin structures of the two types of SDW, it is interesting to discuss the spin correlations in the SC phase. Fig.~\ref{phase}(e) shows 
the subleading SDW interaction in the momentum space. There is a weak peak at $\0q_1$, and a strong one at $\0q_2$. The weaker (stronger) one seems a remnant of that in the C-type (G-type) SDW phase. Indeed, we find in Fig.~\ref{phase}(b) that $\0q_1$ ($\0q_2$) connects pockets of identical (opposite) parity. Although both of them seem to connect gaps of opposite sign, we believe the stronger peak $\0q_2$ here is active for SC, while $\0q_1$ is only passively involved, given the interlayer spin correlations these scattering would cause. This is consistent with the phase diagram, Fig.~\ref{phase}(g), where the singlet SC transition temperature increases until the G-type SDW sets in. Since the G-type SDW vector $\0q_2$ connects $\delta$ and $\gamma$ pockets, we believe that both are important for SC in the \LNO/LAO.
Furthermore, since the density of states of the $z$ orbital increases with $d_{\rm Ni-Ni}$, while that of $x$ orbital almost keeps unchanged, as shown in Fig.~\ref{band_FS}(c), we also conclude that the $3d_{3z^2-r^2}$ orbital is more relevant to the SC in this material. 

We further performed systematic calculations in a window of $d_{\mathrm{Ni-Ni}}$. The results are summarized in the phase diagram, Fig.~\ref{phase}(g). The details of the leading eigenmodes for various values of $d_{\mathrm{Ni-Ni}}$, including those discussed above, are listed in Tables S4 and S5 in the SM\cite{SM} for concreteness.
The ground state changes from G-type SDW to SC and C-type SDW successively with decreasing $d_{\mathrm{Ni-Ni}}$. We propose that this could be achieved by applying a pressure which mainly decreases $d_{\mathrm{Ni-Ni}}$.

{\it Summary and perspective}. 
We have investigated the \LNO \ thin films with interplane distance $d_{\mathrm{Ni-Ni}}$ in a reasonable range. We obtain a C-type SDW, an $s_\pm$-wave SC involving dominant $d_{3z^2-r^2}$ orbitals, and a G-type SDW with increasing $d_{\mathrm{Ni-Ni}}$ and simultaneously increasing transition temperature. The SC is triggered by the G-type SDW fluctuations. The results explain the experimental SC in the thin film under ambient pressure, and also predict tunability of the ground state under pressure. In particular, the experimental verification of the C-type SDW under pressure would shed profound light on the nature of electronic correlations in \LNO, as it is most naturally expected in the itinerant picture as we discussed, but would be difficult in the local-moment picture where interlayer super-exchange is antiferromagnetic.

{\it Acknowledgments}. 
This work is supported by National Key R\&D Program of China (Grants No. 2024YFA1408100, No. 2022YFA1403201), National Natural Science Foundation of China (Grants No. 12074213, No. 12374147, No. 12274205, No. 92365203), and Major Basic Program of Natural Science Foundation of Shandong Province (Grant No. ZR2021ZD01).

Conflict of interest The authors declare that they have no conflict of interest.


\bibliography{327film}

\end{document}